# Efficient low-carbon development in green hydrogen and ammonia economy: a case of Ukraine

**Wadim Strielkowski**
Centre for Energy Studies, Prague Business School, Prague, Czech Republic
e-mail: strielkowski@pbs-education.cz

**Abstract**

This paper focuses on assessing the potentials for the efficient low-carbon development in green hydrogen and ammonia economy using an example of Ukraine as a case study. It describes the country's prerequisites for the transition to renewable energy sources and outlines the ongoing green hydrogen projects. Moreover, it offers a comprehensive SWOT analysis of Ukraine's engagement in the green hydrogen and ammonia economy.

Furthermore, the paper employs a comprehensive bibliometric network analysis using the terms "hydrogen" and "Ukraine" based on a sample of 204 selected publications indexed in the Web of Science (WoS) database. It is doing so by carrying out the network cluster analysis using the text data as well as the bibliometric data with the help of VOSViewer software.

The results and outcomes might be helpful for researchers, stakeholders, and policymakers alike in devising effective strategies and policies aimed at rebuilding and recreating Ukrainian efficient low-carbon energy sector based on the renewable energy sources.

**Keywords:** green energy transition, decarbonization, hydrogen, ammonia, Ukraine

## 1. Introduction

Many post-Soviet economies are facing the issues of economic efficiency and transformation of their energy sectors [1-5]. Sustainable modernization of the power utilities industries represents a cumbersome process with the prospects and risks of renewable energy, such as wind or solar, being constantly assessed and weighted by the relevant stakeholders and policymakers [6-13]. In addition, public acceptance of renewable energy sources remains a problem there, as far as a considerable part of the population still views renewable energy projects as unreliable, utopian, suspicious, and costly [14-19]. The so-called "NIMBY" approach also remains a considerable issue that needs to be overcome [20, 21].

Within this context, Ukraine represents a special case of the efficient low-carbon development. Since 2022, Ukraine's energy infrastructure has suffered extensive damage over the duration of its current war with Russia, creating an urgent need for sustainable energy solutions to support recovery and resilience [22, 23]. Currently, the country is in a painful position in satisfying its short-term energy needs while it is simultaneously embarking on the longer pathway towards the transition towards renewable energy and delivering on its green deal commitments. According to some researchers, renewable energy sources should become the cornerstone of rebuilding the Ukrainian electricity system [24].

One way how to achieve the sustainable development of the energy sector in Ukraine is to build a low-carbon economy with green hydrogen and ammonia becoming the key elements in decarbonizing various sectors such as transport, industry, or power generation.

Due to its location and many favorable geographical factors, Ukraine has excellent potentials to become a major producer of green hydrogen, with its abundance of renewables (particularly wind and solar energy) creating significant opportunities to generate a sustainable fuel source that can boost economic growth and reduce the country's carbon emissions. Shifting in this direction is not only a contribution to global climate obligations, but also a step towards the global green energy economy, stimulating investments and development of clean technologies [25-27]. To effectively support this transformation, Ukraine might want to invest in infrastructure and technology that can increase energy security and create jobs, paving the way for a sustainable economy that emerges as resilient. However, this bold vision would require strong partnerships among government, private enterprises, and research institutions for creating efficient systems and innovative products, optimizing the opportunities for renewable resources and transforming them into large-scale and cost-effective solutions for deploying energy.

## 2. Green hydrogen and ammonia projects in Ukraine

Today, green hydrogen is increasingly becoming the high-capacity energy storage and transport medium of the future that allows countries to open their high renewable energy potential [28-30]. Its potential is especially important in low-and-middle-income countries, such as, for example, Sub-Saharan African (SSA) countries [31-33]. Ukraine is actively pursuing several green hydrogen projects to harness its renewable energy potential and integrate into the European clean energy market.

Green hydrogen in Ukraine is primarily produced through electrolysis, a process that uses electricity generated from renewable sources to split water into hydrogen and oxygen (see Figure 1).

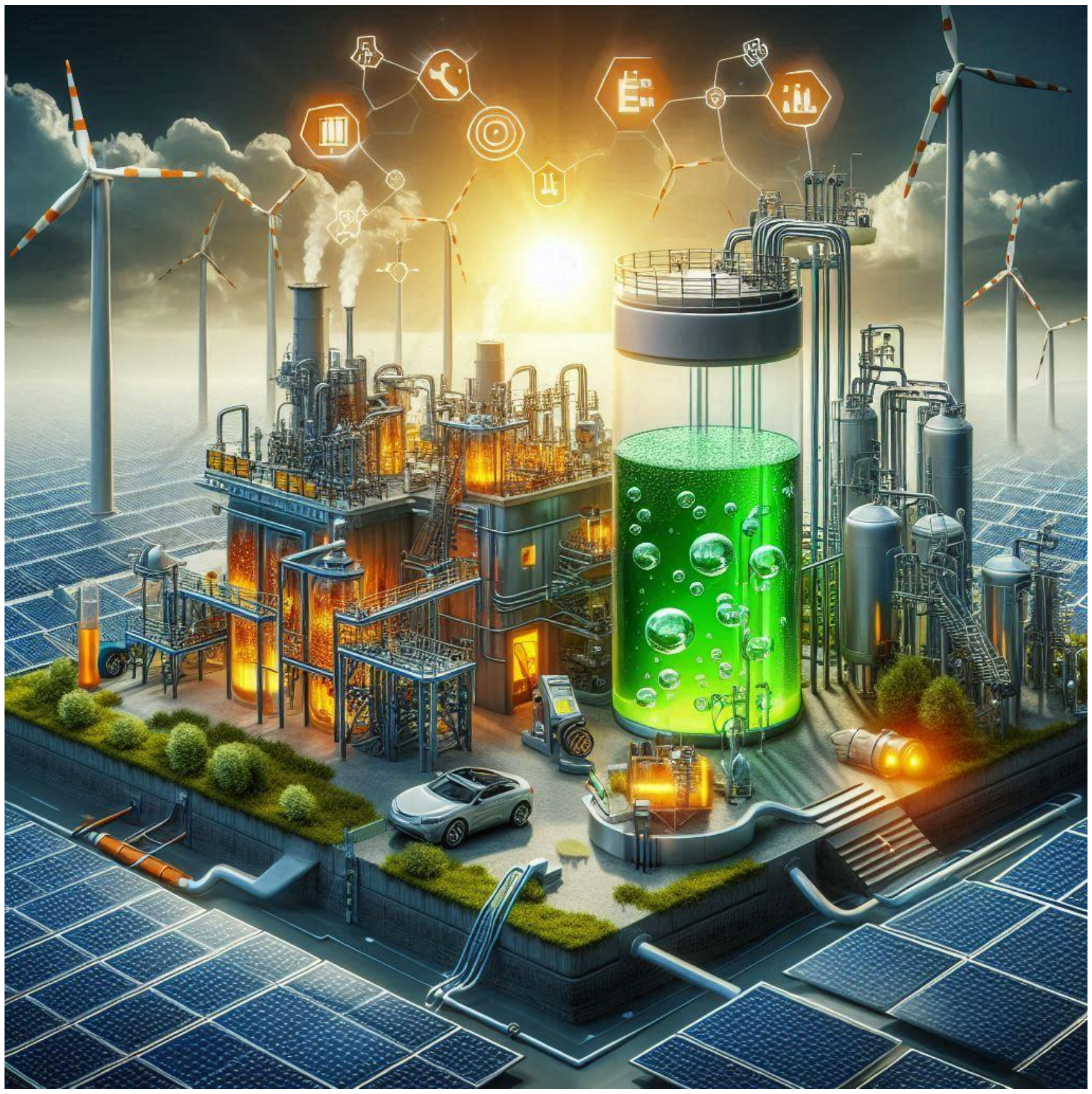

**Fig. 1.** Visualization of green hydrogen production
Source: Own results using Copilot.

Green ammonia, which is produced by combining green hydrogen with nitrogen, is easier to transport than hydrogen and can be used as a hydrogen carrier for international markets [34, 35]. Ammonia is easier to store and transport than hydrogen due to its higher energy density and lower reactivity [35, 36]. When paired with renewable energy, ammonia-derived hydrogen is green hydrogen. Furthermore, ammonia yields a high degree of flexibility, since it can be used directly in fuel cells or converted to hydrogen for industries such as steel or transport. Large-scale renewable energy installations, such as wind farms and solar power plants, provide the electricity needed for electrolysis. For example, the green hydrogen projects in Odesa (southern part of the country) and Zakarpattia (south-western part) rely on these sources. Water for electrolysis is sourced from large water reservoirs and rivers of which Ukraine is abundant. These methods ensure that the hydrogen produced is environmentally friendly and aligns with Ukraine's renewable energy potential.

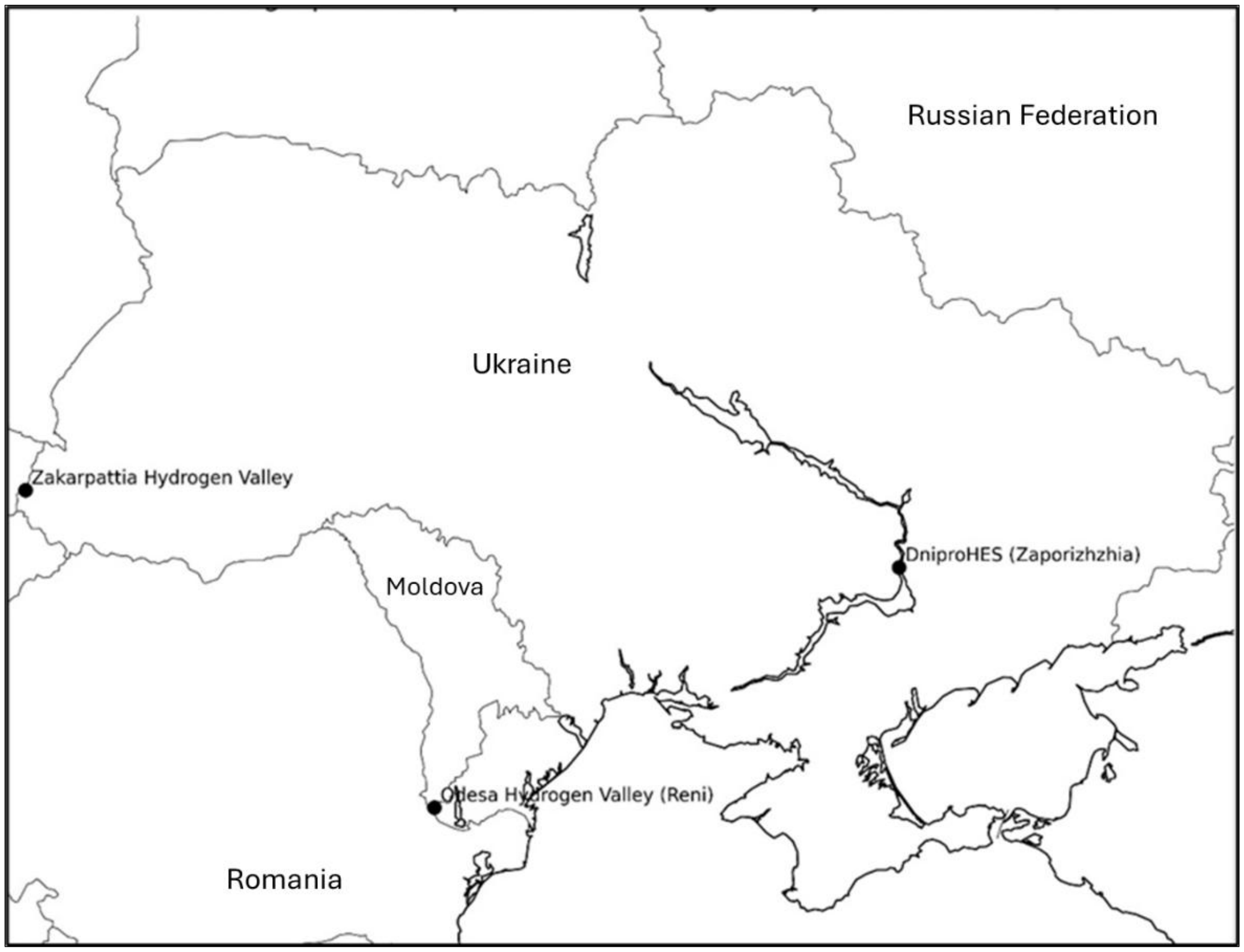

**Fig. 2.** Green hydrogen projects in Ukraine
Source: Own results

The most notable hydrogen energy initiatives and pilot projects in Ukraine include Odesa Hydrogen Valley, Zakarpattia Hydrogen Valley, Central European Hydrogen Corridor, or Ukrhydroenergo's Green Hydrogen Initiative. Figure 2 above locates major green hydrogen projects in Ukraine on the map showing their geographical position within the vast country.

**3. Strengths, weaknesses, opportunities, and threats of hydrogen projects in Ukraine**

There is no doubt that green hydrogen and ammonia projects in Ukraine can help strengthen the country's commitment to developing a green hydrogen economy, leveraging its renewable energy resources, and fostering international collaborations to meet both domestic energy needs and European market demands.

However, alongside the opportunities they might provide for economic development and the transition to the carbon-free economy, their development and deployment is riddled with many obstacles and threats.

The opportunities of the green hydrogen and ammonia projects in Ukraine include the strategic location and renewable resources. Ukraine has abundant wind and solar resources, particularly in southern regions near the Black Sea, facilitating competitive production of green hydrogen and ammonia. In addition, Ukraine's proximity to European markets offers export advantages, aligning with EU ambitions to diversify energy sources away from Russian fossil fuels.

Furthermore, Ukraine can benefit from EU green transition and funding opportunities. European Union's Green Deal and REPowerEU initiative position Ukraine as a potential major exporter of clean hydrogen and ammonia. The access to EU investments, grants, and technological partnerships (including Horizon Europe, EIB financing, EBRD partnerships) can accelerate infrastructure development.

Additionally, Ukraine needs to embark on the path of energy diversification and economic growth. Export-oriented hydrogen and ammonia sectors can attract substantial foreign direct investment (FDI) and create employment and foster economic recovery post-war. The expansion of renewable energy infrastructure linked to hydrogen and ammonia production supports domestic energy resilience and contributes to national decarbonization.

At the same time, Ukraine faces many challenges regarding the development of its green hydrogen and ammonia projects. There is infrastructure damage and rehabilitation. Extensive infrastructure damage from the ongoing war necessitates substantial upfront investments in grid restoration, renewable capacity expansion, and industrial facilities. Persistent instability within the country can deter private investors, elevating reliance on international donor support.

In addition, there is domestic energy security and grid balancing when balancing the demands of domestic electricity supply with energy-intensive hydrogen production requires careful grid management and modernization. The country might face the risks of prioritizing export-focused hydrogen production at the expense of meeting domestic electricity demand, especially during recovery periods.

There also might be a need for increasing and ameliorating Ukraine's technological and workforce capacities. The need for significant technical expertise in green hydrogen production and management, is currently limited due to brain drain and workforce disruptions caused by conflict. Hence, there is a necessity of developing specialized skills, creating educational and training programs, as well as facilitating international collaboration.

The geopolitical risks and impacts on investment of green hydrogen and ammonia projects in Ukraine include continued security threats and investor risk perception. The ongoing geopolitical instability poses major deterrents for long-term foreign investment, with investors wary of asset security, physical protection, and political stability. Elevated insurance costs and difficulty securing financing can undermine the economic viability of large-scale green hydrogen and ammonia projects.

Additionally, Ukraine might face infrastructure vulnerability and military threats. Strategic energy assets, including hydrogen production and export infrastructure, could become military targets, compounding risk perceptions. Thence, vulnerabilities in port infrastructure, rail, and pipeline networks crucial for ammonia export may create bottlenecks.

Ukraine's ambition to become a hydrogen exporter poses the risk of creating new geopolitical dependencies, potentially impacting its national sovereignty and energy policy choices. For example, potential interference or disruption, cyber threats, or sabotage actions could influence investor confidence and operational continuity.

It appears that Ukraine needs to plan how to balance its domestic energy security and export ambitions. The country must develop clear regulations that prioritize domestic energy needs while supporting export capacities, ensuring domestic electricity availability remains stable and affordable. Energy policies should encourage co-development of renewable projects serving dual domestic and export markets, promoting flexibility and resilience.

In order to do that, adopting a phased approach to hydrogen and ammonia sector development, prioritizing domestic renewable expansions needs to be done first, followed by gradual scaling of export capabilities. Avoiding single-market dependencies through diversified export partnerships with EU and non-EU states reduces exposure to geopolitical manipulation or disruptions.

Ukraine might also want to pursue strategic partnerships with international institutions (EU, NATO, G7 nations) to secure investment protection mechanisms, political risk guarantees, and financial support. Diplomatic efforts need to be leveraged to include Ukrainian green hydrogen initiatives within broader security and economic support packages.

Table 1 that follows offers a comprehensive SWOT analysis of the Ukrainian potential for the efficient low-carbon development in green hydrogen and ammonia economy that summarizes the discussion above.

**Table 1:** SWOT analysis

| **Strengths** | **Weaknesses** |
|---|---|
| Abundant wind and solar resources | Extensive infrastructure damage from conflict |
| Strategic geographic proximity to the EU market | Limited current technical expertise in green hydrogen and ammonia sector |
| EU support via Green Deal & REPowerEU initiatives | Ongoing economic strain and funding shortages |
| Existing industrial base suitable for ammonia sector | Potential domestic energy shortages |
| **Opportunities** | **Threats** |
| Attracting significant international investment | Geopolitical instability and continued military threats |
| Positioning Ukraine as a key EU hydrogen and ammonia supplier | Risks of sabotage or cyber-attacks on infrastructure |
| Accelerating economic recovery and growth post-war | Risk of new external dependencies and geopolitical leverage |
| Technological partnerships and access to EU funding | Investor hesitance due to security and ubiquitous corruption concerns |

Source: Own results

## 4. Materials and methods

The empirical part of our study is focused on the bibliometric analysis that is based on a dataset of 204 documents (represented by research papers, conference proceedings, and book chapters) indexed in the Web of Science (WoS) database between 1991 and 2025.

The Web of Science (WoS) database was selected due to being one of the most prestigious and complete abstract and citation databases that exist and offer a vast body of relevant literature on the topic of our research.

Figure 3 that follows offers a Word Cloud diagram highlighting the occurrence and frequency of keywords related to green hydrogen and ammonia projects in Ukraine as well as on hydrogen-related initiatives conducted in Ukraine or by the Ukrainian researchers. The larger and bold keywords demonstrate higher usage frequency.

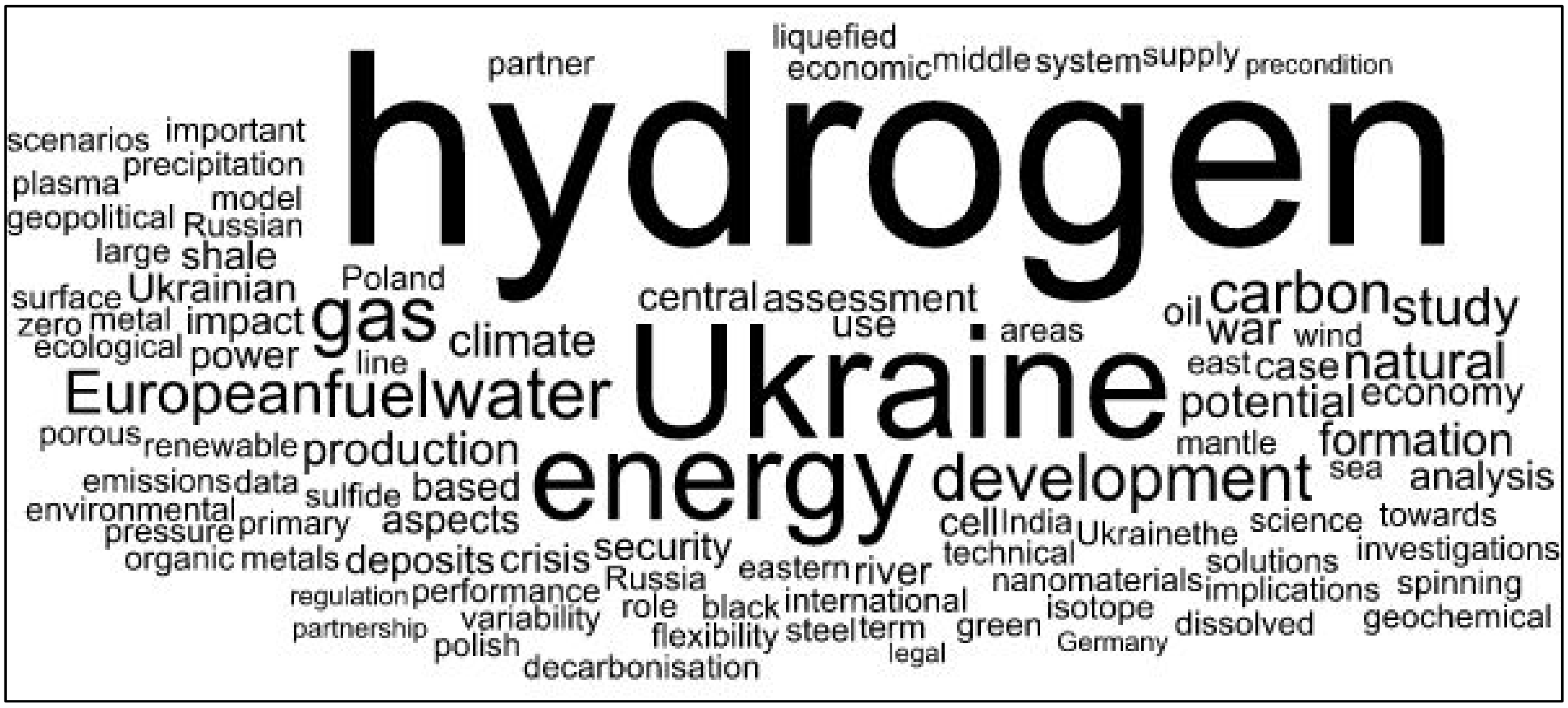

**Fig. 3.** Word Cloud diagram based on 204 selected WoS publications
Source: Own results

Using the set of data described above, aa network cluster analysis of the textual and bibliometric inputs was constructed using VOSViewer software.

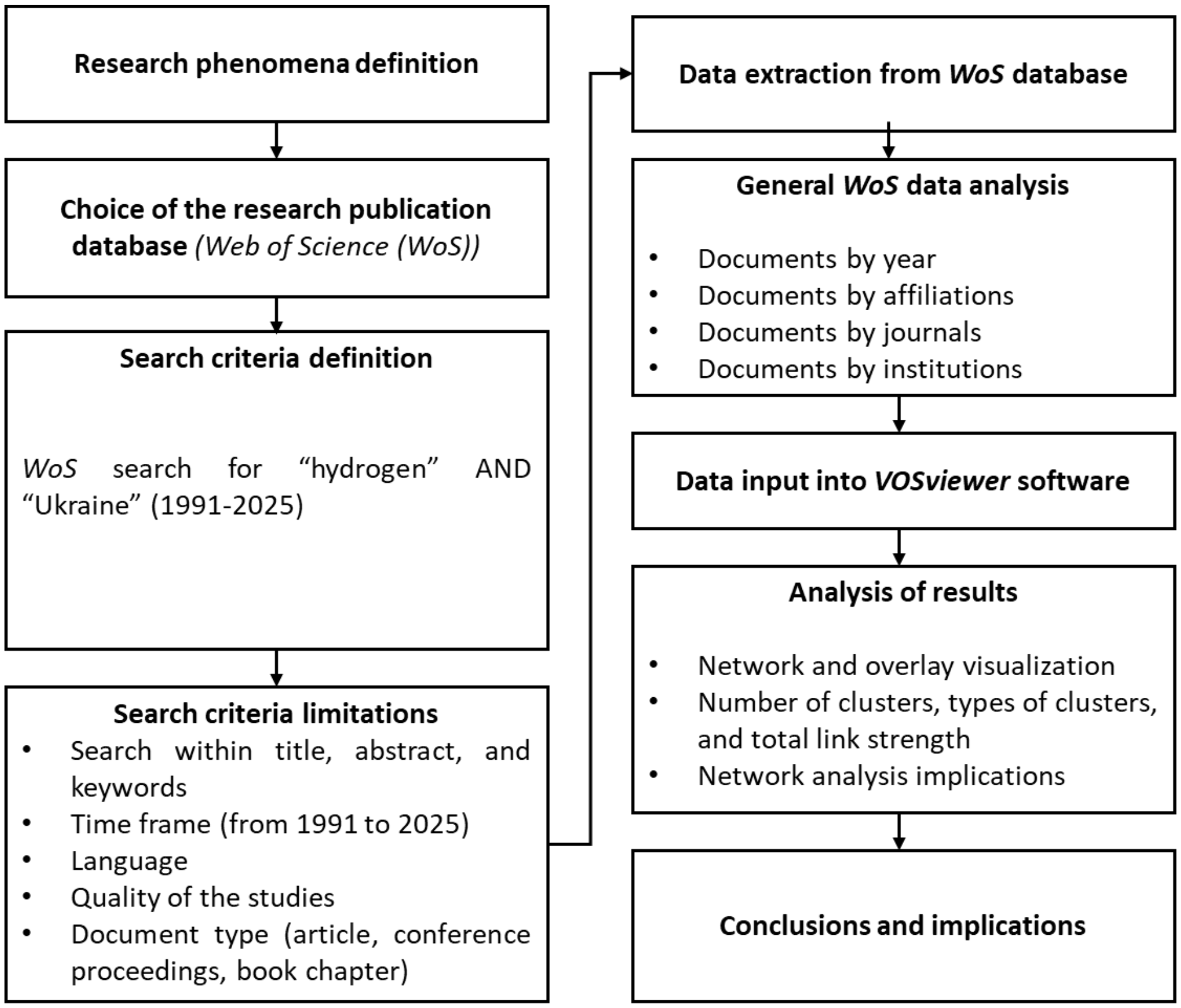

**Fig. 4.** Diagram showing the data selection and network analysis algorithm.
Source: Own results

Figure 4 above provides the description of our algorithm for the data selection, retrieval, processing, as well as the network analysis used in our study. The diagram delineates the methodological approach for data selection and network analysis in a stepwise fashion. Initially, the research phenomenon is defined, which informs the choice of the Web of Science (WoS) as the database for sourcing research publications. The search criteria are then established through the searches in the WoS database for "hydrogen" AND "Ukraine" covering a timeframe from 1991 to 2025.

## 5. Main results

Figure 5 that follows depicts the co-occurrence map based on the text data 204 papers containing the keywords “hydrogen” and “Ukraine” retrieved from WoS database. The analysis identified just two main clusters (marked in red and green).

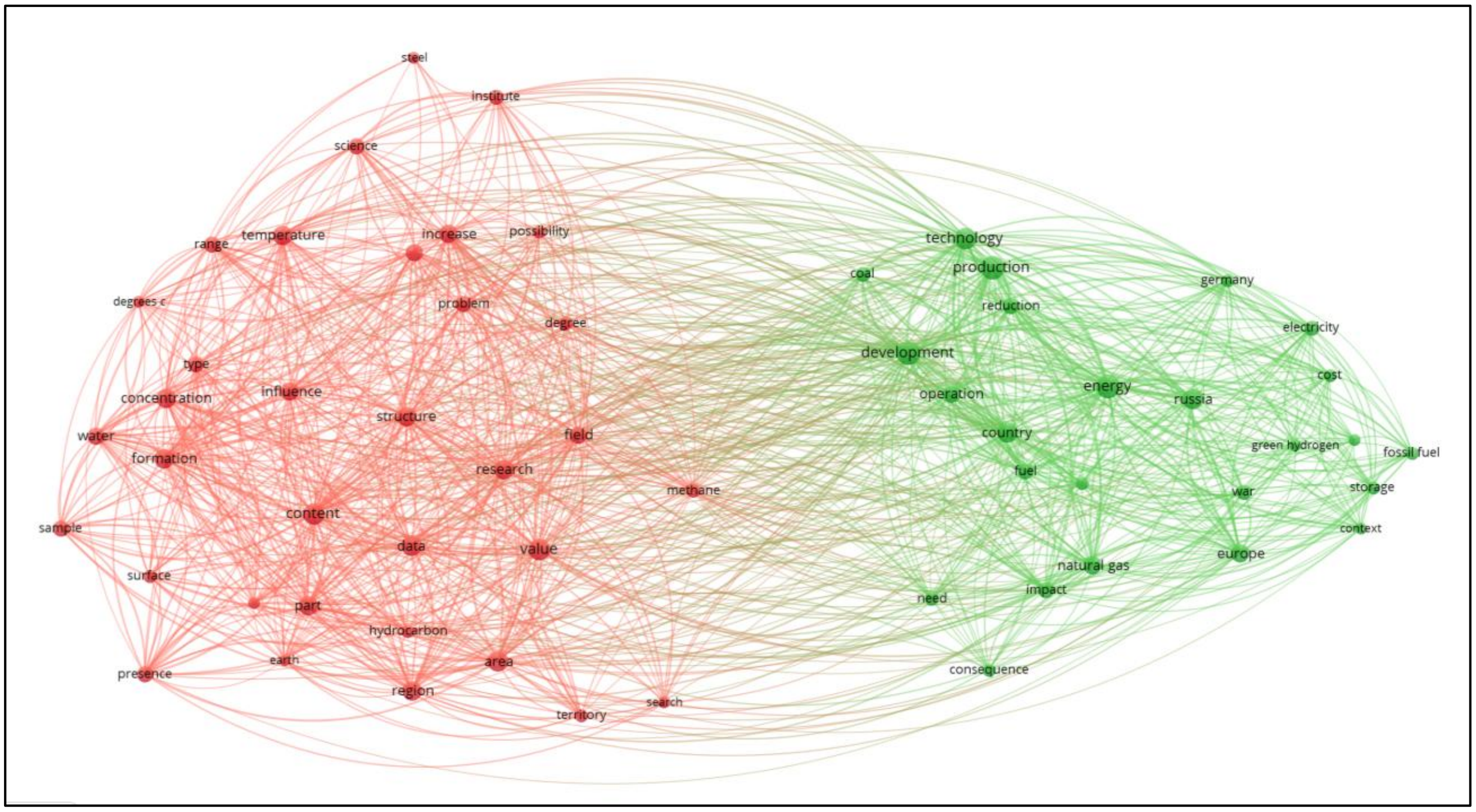

**Fig. 5.** Co-occurrence map based on the text data of the 204 papers containing the keywords “hydrogen” and “Ukraine” retrieved from WoS database
Source: Own results based on VOSViewer v.1.6.19 software

The first cluster, represented in red, primarily encompasses terms related to economic, political, and strategic dimensions, including policy frameworks, geopolitical implications, security issues, and international relations. The second cluster, depicted in green, emphasizes technical, technological, environmental, and developmental themes, particularly focusing on renewable energy integration, green hydrogen production, innovation, and sustainability.

The significant interconnections between the two clusters underscore the interdisciplinary nature of the discourse, reflecting that technological advancements and sustainable hydrogen development are deeply interwoven with geopolitical stability and economic policy considerations. Notably, bridging terms such as "strategy," "energy," "development," and "policy" frequently co-occur across both clusters, highlighting their central role in integrating technical feasibility with strategic planning and policymaking.

This analysis indicates that discussions surrounding Ukraine’s potential in the hydrogen economy are not solely technical or economic but inherently involve geopolitical factors and policy frameworks.

Consequently, Ukraine's successful engagement with the global hydrogen market will necessitate a comprehensive approach that effectively integrates technological capabilities, sustainable development strategies, and robust geopolitical and economic policy frameworks.

Furthermore, the bibliographic map, generated from bibliometric analysis of 204 scientific publications retrieved from the Web of Science database containing the keywords "hydrogen" and "Ukraine" (see Figure 6 below) illustrates several key thematic clusters based on keyword co-occurrence. The visualization reveals four primary clusters, each distinguished by color and representing unique thematic groupings within the scholarly literature.

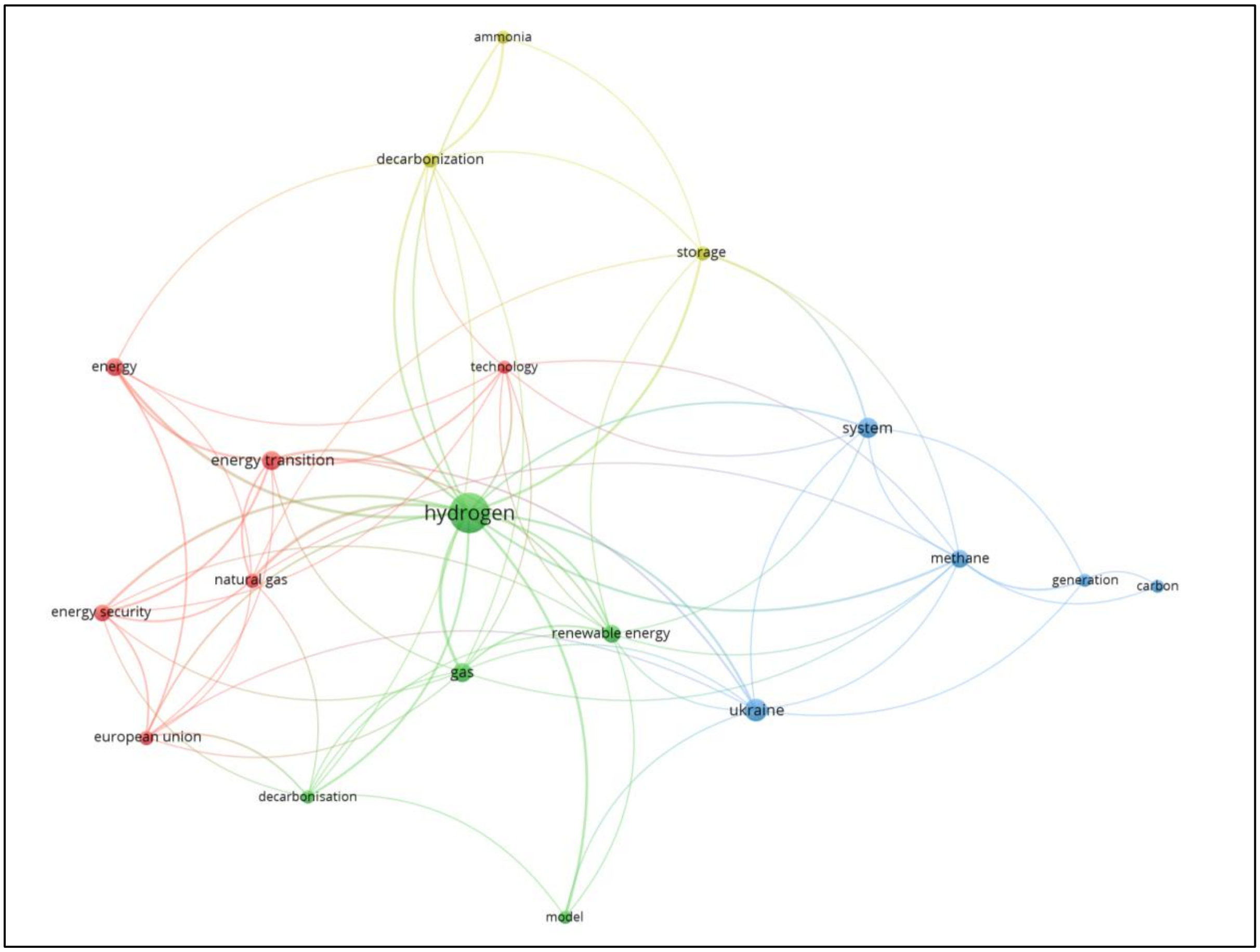

**Fig. 6.** Bibliographic map based on the bibliometric data (co-authorship, keyword co-occurrence, citation, bibliographic coupling, or co-citation map) from 204 papers containing the keywords “hydrogen” and “Ukraine” in WoS database.
Source: Own results based on VOSViewer v.1.6.19 software

The central green cluster prominently features the term "hydrogen" and closely associated terms such as "renewable energy," "solar," and "decarbonisation," highlighting a strong research focus on sustainable and renewable energy aspects of hydrogen technologies. This cluster indicates that current research extensively explores renewable energy integration and decarbonization pathways associated with hydrogen in Ukraine. The red cluster emphasizes energy transition and security aspects, including keywords such as "energy transition," "energy security," "natural gas," and "European Union." This suggests a significant body of literature addressing hydrogen's strategic role in the broader context of energy policy, geopolitics, and European energy security frameworks.

The yellow cluster connects "ammonia," "storage," "technology," and "decarbonization," emphasizing technological development, storage solutions, and ammonia production pathways as vital elements of hydrogen research in Ukraine. This cluster underscores the critical role of technological advancement and infrastructure development within the hydrogen economy.
Lastly, the blue cluster includes terms such as "system," "methane," "generation," "carbon," and "Ukraine," reflecting studies focused on hydrogen production processes, system integration, carbon management, and the technical aspects related to hydrogen infrastructure and generation methods specific to Ukraine.
Overall, the bibliometric map demonstrates an interdisciplinary research landscape, highlighting the integration of technological advancements, renewable energy adoption, geopolitical considerations, and strategic planning. These findings suggest that the successful advancement of Ukraine's hydrogen sector requires collaborative efforts across technological, economic, environmental, and geopolitical domains.

**6. Conclusions**

To sum this all up, while geopolitical and security risks pose substantial challenges, Ukraine's renewable energy potential, strategic location, and European partnership opportunities present significant prospects for the creation of the efficient green hydrogen and ammonia economy. Balancing domestic energy security with ambitious export goals through strategic policies, phased implementation, and risk mitigation strategies will be essential for successful engagement in the green hydrogen and ammonia economy.
In order to achieve progress in creating an efficient low-carbon development in green hydrogen and ammonia economy, Ukraine needs to clearly articulate a national green hydrogen strategy linked to EU alignment and post-war recovery goals, signaling stability to investors. Rebuilding resilient infrastructure (renewable energy plants, storage facilities, transportation networks) under international cooperation frameworks needs prioritized and supported with preferential projects, grants, and loans via state and private sector support.
It becomes apparent that to successfully develop its green hydrogen and ammonia project for rebuilding and transforming its low-carbon sustainable energy sector, Ukraine needs to seek comprehensive insurance schemes and risk guarantees to mitigate geopolitical risk for international investors. To do that, the country needs to foster local technical expertise, education, and job creation to ensure long-term viability of the hydrogen and ammonia economy.